\documentclass[fleqn,10pt,twocolumn]{SICE_ISCS}

\usepackage{epsfig}    
\usepackage{amsmath}   
\usepackage{amssymb}   

\usepackage{amsthm} 

\usepackage{bm}            
\usepackage{algorithm}     
\usepackage{algpseudocode} 

\theoremstyle{definition}
\newtheorem{mydef}{Definition}
\newtheorem{myprop}{Property}
\newtheorem{myproblem}{Problem}
\newtheorem{myrem}{Remark}

\title{Finding a Feasible Initial Solution for Flatness-Based Multi-Link Manipulator Motion Planning under State and Control Constraints}

\author{Keisuke Uto${}^{1\dagger}$, Makoto Obayashi${}^{1}$ and Gaku Takano${}^{1}$}
\speaker{Keisuke Uto}

\affils{${}^{1}$Denso IT Laboratory, Japan\\
(E-mail: \{kuto, mohbayashi, gtakano\}@d-itlab.co.jp)\\
}

\abstract{
In this paper, we present a method to initialize at a feasible point and unfailingly solve a non-convex optimization problem in which a set-point motion is planned for a multi-link manipulator under state and control constraints.
We construct an initial feasible solution by analyzing the final time effect for feasibility problems of flatness based motion planning problems.
More specifically, we first find a feasible time-optimal trajectory under state constraints without a control constraint by solving a linear programming problem. Then, we find a feasible trajectory under control constraints by scaling the trajectory.
To evaluate the practical applicability of the proposed method, we did numerical experiments to solve a multi-link manipulator motion planning problem by combining the method with recursive inverse dynamics algorithms.
}

\keywords{%
Optimal Control, Mechanical Systems Control, Control Applications
}
\begin{document}
\maketitle
\thispagestyle{empty}
\pagestyle{empty}

\section{Introduction}
To make a highly automated system, motion planning is more important than following a reference trajectory given by humans.
To plan the motion of robotic manipulators, the motion can be ineffective and sometimes physically impossible if we inexactly deal with nonlinear dynamics of the manipulators and constraints like joint angle and torque limit.
In this paper, we tackle a continuous-time set-point motion planning problem of multi-link manipulators under state and control constraints by using nonlinear dynamics exactly.

A typical method for motion planning that consider dynamics under some constraints discretizes state equations by using small time grids and solves their constrained optimization problems (e.g. Multiple Shooting Methods).
However, such methods suffer from huge numbers of variables and numerical integration errors when the methods are used to try to solve long time horizon problems.
We adopt flatness based motion planning methods because they can formulate long time horizon problems that have few parameters while considering nonlinear dynamics precisely and avoiding numerical integration errors.

In planning using a flatness based method under control constraints, its optimization problem generally becomes non-convex due to nonlinear dynamics.
Therefore, if we solve the problem with a general nonlinear programming (NLP) algorithm like sequential quadratic programming (SQP) or interior point methods, the behavior of the algorithm varies depending on the initialization. Thus, there is no guarantee that a feasible solution can be found even if the problem has feasible regions.
\cite{ibrahim2008improving} \cite{chinneck2007feasibility}. (In our numerical experiments of manipulator planning problems, NLP algorithms often failed to find a feasible solution when we tried to use several initial points.)

We aim to reveal details of the optimization problem structure of the flatness based motion planning algorithm under state and control constraints and construct a method that can unfailingly find feasible solutions of it.

We derive a method to initialize a solution algorithm with a feasible initial point by parameterizing a trajectory with a linear combination of basis functions and by time transformation.
Our contributions are summarized as follows.

\begin{itemize}
\item By combining time scaling transformation with parameterization of a linear combination of an arbitrary basis function, we convert the planning problem to analyze the effect of final time on the optimization problem structure.
\item We describe a time-optimal feasible solution under state constraints that uses linear programming(LP) and use it to construct a method to find  feasible initial points under control constraints.
\item We present details of practical implementations to solve multi-link manipulator planning problems with numerical simulation by combining our initialization method and recursive inverse dynamics algorithms.
\end{itemize}
\section{Related Work}
A method that deals with a state equation constraint by locating sample times in the control horizon and calculating the states by numerical integration is called a ``shooting method'' \cite{betts2010practical}.
There are many types of research that extend this family of methods and reduce their computational cost.
We focus on a method that does not use numerical integration of a state equation because a state equation of a Multi-Link robot manipulator is difficult to numerically integrate both very accurately and quickly.

The differential flatness concept was proposed by Fliess et al.\cite{fliess1995flatness}. Flatness has been studied from the viewpoint of motion planning problems \cite{martin2003flat}.
Louembet et al.\cite{louembet2010motion} developed a convex programming algorithm for a motion planning problem under semi-infinite inequality constraints by approximating  a feasible region to a convex polytope.
Van Looke et al.\cite{van2015optimal} developed a method to approximate polynomial semi-infinite inequality constraints by using the convex hull property of B-splines \cite{de1978practical}.
Furthermore, they applied the methods to a flexible robot arm planning problem \cite{van2015b}.
We follow this line of research further and tackle more practical problem settings on the basis of more complex dynamics.

In robotics, Martin and Bobrow \cite{martin1999minimum}, Wang et al.\cite{wang2001payload}, and Lee et al.\cite{lee2005newton} have solved motion planning problems of multi-joint robot manipulators by parameterizing joint trajectories with B-splines, though they did not use the flatness concept explicitly.
The theoretical basis of these methods becomes clearer from the viewpoint of flatness with our formulation, and their formulation can be basically understood to be the same as our formulation of flatness-based motion planning.

Their methods are based on the Lie group formulation of multi-link robot manipulator modeling and effective calculation of the inverse dynamics of a manipulator.
The theory of the Lie group formulation of robot dynamics is described in detail elsewhere \cite{murray1994mathematical}, \cite{park1995lie}.
A recursive algorithm for the Lie group formulation of robot dynamics was developed \cite{park1995lie} and then analyzed more deeply and extended \cite{ploen1999coordinate} \cite{sohl2001recursive}.
Our method references Kim and Polland's implementation \cite{kim2011fast}.

Wang et al. \cite{wang2001payload} solved a problem by formulating control constraints as costs, which are soft constraints.
We deal with control constraints as hard constraints to avoid the weight of control constraints, which vary depending on cost weight tuning.
\section{Problem Formulation}
In this section, we provide a basic formulation of a flatness based motion planning of robotic manipulators.
In this paper, all vectors are column vectors and $ {\top} $ means vector or matrix transposition.

\subsection{Flatness Based Motion Planning of Robot Manipulator}
A nonlinear system $ \dot{\bm{x}}(t) = \bm{f}(\bm{x}(t),\bm{u}(t)) $ is differentially flat or flat if there exists one-to-one mapping from a tuple of state and control $ (\bm{x}(t),\bm{u}(t)) $ to flat outputs $ \bm{y}(t) $ that satisfy the below equations.
%
\begin{mydef}[Differential Flatness \cite{fliess1995flatness}]
\begin{align}
\bm{y}(t) &= \bm{\phi} (\bm{x}(t), \bm{u}(t), \dot{\bm{u}}(t), \cdots, \bm{u}^{(\alpha)}(t))                  \label{flat_y} \\
\bm{x}(t) &= \bm{\psi}_{\bm{x}} (\bm{y}(t), \dot{\bm{y}}(t), \ddot{\bm{y}}(t), \cdots, \bm{y}^{(\beta)}(t))    \label{flat_x} \\
\bm{u}(t) &= \bm{\psi}_{\bm{u}} (\bm{y}(t), \dot{\bm{y}}(t), \ddot{\bm{y}}(t), \cdots, \bm{y}^{(\beta+1)}(t))  \label{flat_u}
\end{align}
\end{mydef}
%
The equation of the motion of a mechanical system, like that of a robot manipulator, is generally formulated as follows \cite{murray1994mathematical}.
(The friction term is omitted for the sake of simplicity).
%
\begin{equation}
\bm{\tau} = M(\bm{q})\ddot{\bm{q}} + C(\bm{q},\dot{\bm{q}})\dot{\bm{q}} + \bm{G}(\bm{q}) \label{eqn_roboeom}
\end{equation}
where
$ \bm{q}(t) \in {\mathbb{R}}^n $ are joint trajectories,
$ M(\bm{q}) \in {\mathbb{R}}^{n \times n} $ is a generalized inertia matrix, 
$ C(\bm{q},\dot{\bm{q}}) \in {\mathbb{R}}^{n \times n} $ is a Coriolis matrix, 
$ \bm{G}(\bm{q}) \in {\mathbb{R}}^n $ is a potential term,
and $ \bm{\tau} \in {\mathbb{R}}^n $ are torque control inputs.
Robot manipulators are flat systems if we regard joint trajectories $ \bm{q}(t) $ as flat outputs.
In other words, $ \bm{q}(t), \dot{\bm{q}}(t), \ddot{\bm{q}}(t) $ and $ \bm{\tau}(t) $ can be derived only by the $ \bm{q}(t) $ trajectory.
Therefore, we can plan both state and control trajectories of a manipulator by solving a function optimization problem only of a joint trajectory $ \bm{q}(t) $ that is second order differentiable.
In this paper, we focus on the following a set-point problem under linear state and control constraints. 
%
\begin{myproblem}[Flatness Based Robot Motion Planning Problem] \label{problem_flat_t}
\begin{align}
& \underset{\bm{q}(\cdot), t_f}{\text{min}} \quad J(\bm{q}(\cdot),t_f) \notag \\
& s.t. \notag \\
& \quad \bm{q}(0) = \bm{q}_0,        \quad  \bm{q}(t_f) = \bm{q}_f \notag \\
& \quad \dot{\bm{q}}(0) = \dot{\bm{q}}_0,  \quad  \dot{\bm{q}}(t_f) = \dot{\bm{q}}_f \notag \\
& \quad \underline{\bm{q}} \le \bm{q}(t) \le \overline{\bm{q}} \notag \\
& \quad \underline{\dot{\bm{q}}} \le \dot{\bm{q}}(t) \le \overline{\dot{\bm{q}}} \notag \\
& \quad \underline{\bm{\tau}} \le \bm{\tau}(\bm{q}(t),\dot{\bm{q}}(t),\ddot{\bm{q}}(t)) \le \overline{\bm{\tau}} \qquad (t \in [0,t_f] ) \notag \\
& where \notag \\
& \quad \bm{\tau}(\bm{q},\dot{\bm{q}},\ddot{\bm{q}}) = M(\bm{q})\ddot{\bm{q}} + C(\bm{q},\dot{\bm{q}})\dot{\bm{q}} + \bm{G}(\bm{q}) \label{eqn_flat_opt_robot}
\end{align}
\end{myproblem}
$ \underline{\bm{q}}, \overline{\bm{q}}, \underline{\dot{\bm{q}}}, \overline{\dot{\bm{q}}} \in {\mathbb{R}}^{n},
  \quad \underline{\bm{\tau}}, \overline{\bm{\tau}} \in {\mathbb{R}}^{n} $
 are lower or upper bounds constants with respect to joint angles, joint angular velocities, and input torques of each joint.
The cost function $ J $ is specified in accordance with task settings.
$ t_f \in {\mathbb{R}_{+}} $ is the final time of the control horizon, 
We formulate both free and fixed final time problems in a unified way by assuming $ t_f $ of {\bf Problem \ref{problem_flat_t}} as a variable or a constant.
%
\subsection{Time Scaling Transformation}
To analyze our optimization problem structure of both free and fixed final time and separate parameterization of a trajectory and final time,
we normalize control time horizon $ t \in [0,t_f] $ to $ s \in [0,1] $ by time scaling transformation.
Then our trajectories of joint angle, angular velocity, angular acceleration, and torque are formalized as follows.
\begin{align}
\bm{q}(t)        &= \tilde{\bm{q}}(s) \\
\dot{\bm{q}}(t)  &= \tilde{\bm{q}}^{\prime}(s) \dot{s} \\
\ddot{\bm{q}}(t) &= \tilde{\bm{q}}^{\prime}(s) \ddot{s} + \tilde{\bm{q}}^{\prime\prime}(s) \dot{s}^2 \\
\tilde{\bm{\tau}}(s)  &= M(\tilde{\bm{q}})\tilde{\bm{q}}^{\prime}\ddot{s}
  + [M(\tilde{\bm{q}})\tilde{\bm{q}}^{\prime\prime} + C(\tilde{\bm{q}},\tilde{\bm{q}}^{\prime})\tilde{\bm{q}}^{\prime}]\dot{s}^2
  + \bm{G}(\tilde{\bm{q}})
\end{align}
In this paper, we write the time normalized function of variable $ s $ as a tilde over a character like ``$ \;\tilde{}\; $'' and ``\;$ '\; $'' for derivatives of $ s $ . 
\subsection{Trajectory Parameterization with Linear Combination of Basis Functions}
We parameterize $ \tilde{q}_i(s) $ as a linear combination of $ m $ basis functions $ b_j(s) $ as $ \tilde{q}_i(s) = \sum_{j=1}^m a_{ij} b_j(s) $ and $ s(t) $ as linear scaling of $ t $ to represent a large class of trajectory and to parameterize simply.
Then, we can derive the following trajectory formulation.
\begin{align}
s        &= t_f^{-1} t \quad (s \in [0,1]) \notag \\
\dot{s}  &= t_f^{-1} \notag \\
\ddot{s} &= 0 \notag \\
\tilde{\bm{q}}(s)              &= \sum_{j=1}^m \bm{a}_{j} b_{j}(s) = A \bm{b}(s) \\
\tilde{\bm{q}}^{\prime}(s)      &= A \bm{b}^{\prime}(s) \\
\tilde{\bm{q}}^{\prime\prime}(s) &= A \bm{b}^{\prime\prime}(s) \\
\tilde{\bm{\tau}}(s) &=
   [M(\tilde{\bm{q}})\tilde{\bm{q}}^{\prime\prime} + C(\tilde{\bm{q}},\tilde{\bm{q}}^{\prime})\tilde{\bm{q}}^{\prime}] t_f^{-2} + \bm{G}(\tilde{\bm{q}})
\end{align}
where $ \bm{a}_{j} \in {\mathbb{R}}^{n} $, $ A = [\bm{a}_{1}, \bm{a}_{2}, \cdots , \bm{a}_{m}] \in {\mathbb{R}}^{n \times m} $, and $ \bm{b}(s) \in {\mathbb{R}}^{m} $ .
Time scaled and trajectory parameterized {\bf Problem \ref{problem_flat_t}} is formulated as follows.
\begin{myproblem}[Time Scaled Motion Planning Problem] \label{problem_timescaled}
\begin{align}
& \underset{A \in {\mathbb{R}}^{n \times m}, t_f \in {\mathbb{R}}}{\text{min}} \quad J(A,t_f) \notag \\
& s.t. \notag \\
& \quad A \bm{b}(0) = \bm{q}_0,                   \quad  A \bm{b}(1) = \bm{q}_f \notag \\
& \quad A \bm{b}^{\prime}(0) = \dot{\bm{q}}_0 t_f, \quad  A \bm{b}^{\prime}(1) = \dot{\bm{q}}_f t_f \notag \\
& \quad \underline{\bm{q}}       \le A \bm{b}(s)         \le \overline{\bm{q}} \notag \\
& \quad t_f \underline{\dot{\bm{q}}} \le A \bm{b}^{\prime}(s) \le t_f \overline{\dot{\bm{q}}} \notag \\
& \quad \underline{\bm{\tau}} \le \tilde{\bm{\tau}}(s; A,t_f) \le \overline{\bm{\tau}} \quad (s \in [0,1] ) \notag \\
& \quad t_f \ge 0 \notag \\
& where \notag \\
& \quad \tilde{\bm{\tau}}(s; A,t_f) =    [M(\tilde{\bm{q}})\tilde{\bm{q}}^{\prime\prime} + C(\tilde{\bm{q}},\tilde{\bm{q}}^{\prime})\tilde{\bm{q}}^{\prime}] t_f^{-2} + \bm{G}(\tilde{\bm{q}}) \label{eqn_main_problem}
\end{align}
\end{myproblem}
The main challenge in this paper is this formulation of a problem.
{\bf Problem \ref{problem_timescaled}} has the following difficulty.
\begin{itemize}
\item Inequality constraints are semi-infinite constraints on a continuous interval of $ s $.
\item Torque $ \tilde{\bm{\tau}}(s; A,t_f) $ constraint is generally nonlinear and non-convex.
\end{itemize}
We can relax the former problem by evaluating finite sample points on $ s_i \in [0,1] $ .
The latter non-convex problem is the fundamental difficulty of our optimization problem, and in our numerical experiments, we often failed to solve the problem by directory applying a standard nonlinear programming(NLP) algorithm like SQP and interior point methods.

{\bf Problem \ref{problem_timescaled}} has no obvious feasible points and if we solve the problem by using NLP with a infeasible initial point, the algorithm may stick at an infeasible point near it, and there is no guarantee to find a feasible point even if the problem has a feasible region.
\section{Initialization Method with a Feasible Initial Point}
To unfailingly find a solution for our non-convex {\bf Problem \ref{problem_timescaled}} in feasible settings, we think that one reliable solution is initializing with a feasible point.
To construct the feasible point, first we solve a convex sub-problem that satisfies state constraints and then modify its solution to satisfy control constraints.
Our method is composed of three steps and overviewed as {\bf Algorithm \ref{alg_feas_init}}.
We next describe details of the algorithm and how the algorithm can find a feasible point.

\begin{algorithm}
\caption{Feasible Initialization and Improving}
\label{alg_feas_init}
\begin{algorithmic}
\renewcommand{\algorithmicrequire}{\textbf{Input:}}
\renewcommand{\algorithmicensure}{\textbf{Output:}}
\Require  $ \bm{q}_0, \dot{\bm{q}}_0, \bm{q}_f, \dot{\bm{q}}_f, \underline{\bm{q}}, \overline{\bm{q}}, \underline{\dot{\bm{q}}}, \overline{\dot{\bm{q}}}, \underline{\bm{\tau}}, \overline{\bm{\tau}} \in {\mathbb{R}}^{n} $
\Ensure $ A^{opt} \in {\mathbb{R}}^{n \times m} , t_f^{opt} \in {\mathbb{R}} $
\State $ \{A^{LP}, t_f^{LP}\} \leftarrow $ Solve {\bf Problem \ref{problem_LP}} with LP()
\State \Comment{Step1: State Constraint Satisfaction}
\State \hspace{2.5em} $ t_f^{feas} \leftarrow $ LineSearch($ t_f^{LP} $)
\State \Comment{Step2: State \& Control Constraint Satisfaction}
\State $ \{A^{opt}, t_f^{opt}\} \leftarrow $ NLP($ A^{LP}, t_f^{feas}$)
\State \Comment{Step3: Feasible Solution Improvement}
\State \textbf{return} $ \{A^{opt}, t_f^{opt}\} $
\end{algorithmic}
\end{algorithm}

%
\subsection{Time-Optimal Feasible Point under State Constraints}
We first ignore torque constraints and solve the time-optimal trajectory under state constraints that are relaxed to evaluate finite sample points in control intervals.
This sub-problem can be effectively solved with linear programming(LP).
\begin{myproblem}[Time-Optimal Motion Planning Problem under State Constraints (Linear Programming)] \label{problem_LP}
\begin{align}
& \{A^{LP},t_f^{LP}\} = \underset{A \in {\mathbb{R}}^{n \times m}, t_f \in {\mathbb{R}}}{\text{argmin}} \quad t_f \notag \\
& s.t. \notag \\
& \quad A \bm{b}(0) = \bm{q}_0,                   \quad  A \bm{b}(1) = \bm{q}_f \notag \\
& \quad A \bm{b}^{\prime}(0) = \dot{\bm{q}}_0 t_f, \quad  A \bm{b}^{\prime}(1) = \dot{\bm{q}}_f t_f \notag \\
& \quad \underline{\bm{q}}       \le A \bm{b}(s_i)         \le \overline{\bm{q}} \label{ineq_q} \\
& \quad t_f \underline{\dot{\bm{q}}} \le A \bm{b}^{\prime}(s_i) \le t_f \overline{\dot{\bm{q}}} \quad (i \in \{1,2,...,N\}) \label{ineq_qdot} \\
& \quad t_f \ge 0 \label{eqn_LP}
\end{align}
\end{myproblem}
%
\subsection{Feasible Initial Point under State and Control Constraints}
Next, we construct a feasible initial point under control constraints without violating state constraints using $ \{A^{LP},t_f^{LP}\} $.
We use the following property.
\begin{myprop}[Relaxed Torque Constraints Check] \label{prop_torque}

If $ \dot{\bm{q}}_0 $ and $ \dot{\bm{q}}_f = 0 $
and whole trajectory of the solution of {\bf Problem \ref{problem_LP}}: $ \tilde{\bm{q}}(s) = A^{LP} \tilde{\bm{b}}(s) $ satisfies
$ \underline{\bm{\tau}} < \bm{G}(A^{LP} \bm{b}(s)) < \overline{\bm{\tau}} $ ,
there exists $ t_f >= t_f^{LP} $ that satisfies torque constraints.
\end{myprop}
%
\begin{proof}
Using torque expression in (\ref{eqn_main_problem})
\[
\lim_{t_f \to \infty} \tilde{\bm{\tau}}(s; A,t_f) = \bm{G}(A \bm{b}(s))
\]
Therefore, if $ t_f $ is large, the torque constraints are 
$ \underline{\bm{\tau}} \le \epsilon + \bm{G}(A \bm{b}(s)) \le \overline{\bm{\tau}} $ for $ |\epsilon| \ll 1 $ . 
This inequality is satisfied if $ \underline{\bm{\tau}} < \bm{G}(A^{LP} \bm{b}(s)) < \overline{\bm{\tau}} $ is always satisfied for $ \forall{s} \in [0,1] $ .
If we make $ t_f $ larger, the range of constraints $ t_f \underline{\dot{\bm{q}}} \le A \bm{b}^{\prime}(s_i) \le t_f \overline{\dot{\bm{q}}} $ expands and is not violated.
\qedhere 
\end{proof} 
Practically, we find suitable $ t_f^{feas} $ by a line search in the second step of the algorithm.
In the last step, we apply a NLP algorithm using this feasible initial solution $ \{A^{LP},t_f^{feas}\} $ and improve the solution.
\begin{myrem}[About Assumption in {\bf Property \ref{prop_torque}}]

Assumption $ \dot{\bm{q}}_0 $ and $ \dot{\bm{q}}_f = 0 $ is satisfied if the motion is from a stop state to a stop state, which is the case in a practical situation.
Assumption $ \underline{\bm{\tau}} < \bm{G}(A^{LP} \bm{b}(s)) < \overline{\bm{\tau}} $ is always satisfied in practical task settings for an industrial manipulator.
Thus, we think these assumptions cover a large class of practical applications. A general case is an interesting problem but is future work.
\end{myrem}
%
We summarize {\bf Algorithm \ref{alg_feas_init}} and call the solutions of each step in the algorithm as follows.
\begin{itemize}
\item {\bf solution LP:} $ \bm{q}(t)^{LP}   = A^{LP} \bm{b}((t_f^{LP})^{-1} t) $ \\
 a time-optimal solution that satisfies state constraints solved with LP
\item {\bf solution feas:}\footnote{{\bf Solution feas} is made by expanding {\bf solution LP}, thus variable A is not changed. } $ \bm{q}(t)^{feas} = A^{LP} \bm{b}((t_f^{feas})^{-1} t) $ \\
 a solution that satisfies both state and control constraints
\item {\bf solution opt:} $ \bm{q}(t)^{opt}  = A^{opt} \bm{b}((t_f^{opt})^{-1} t) $ \\
 a solution {\bf solution feas} improved with NLP by minimizing cost in accordance with task settings
\end{itemize}
We have described the main procedure of our algorithm. 
However, to implement our algorithm for practical multi-link manipulators, inequality constraints need to be efficiently checked.
We next describe the techniques to do this.

\section{Practical Implementation Techniques to Calculate Torque of Multi-Link Manipulators}
To check torque constraints at normalized time $ s_i $, we have to calculate $ \tilde{\bm{\tau}}(\tilde{\bm{q}}(s_i), \tilde{\bm{q}}^{\prime}(s_i), \tilde{\bm{q}}^{\prime\prime}(s_i), t_f) $ by using an inverse dynamics algorithm.
The problem here is the very high computational cost to calculate inverse dynamics for a multi-link robot manipulator such as a 6-degrees of freedom (DoF) one because a motion equation of the manipulator is extremely complex and almost impossible to write down in closed form explicitly.
To calculate torque efficiently, we use a recursive inverse dynamics algorithm formulated by a Lie group.
\subsection{Lie Group Formulation of Recursive Inverse Dynamics}
An equation of motion of a multi-link robot manipulator obtains a high-level, coordinate-free view when it is formulated in a Lie group \cite{murray1994mathematical}, \cite{park1995lie}.
We use a Lie group formulation of robot dynamics because it is useful to analyze and improve from a geometric viewpoint.
An order of the computational cost of calculating inverse dynamics by using a closed-form equation of motion is $ O(n^4) $, where $ n $ is the number of joints, which is very large \cite{featherstone2000robot}\cite{ploen1999coordinate}.
We use a recursive calculation algorithm for this, which is known to have a low computational cost at order $ O(n) $ . 

In the recursive algorithm, an equation of motion is divided into recursive equations by introducing intermediate variables.
First, the algorithm computes states of joints from a base of a manipulator to the end-effector by forward recursion. Then, torque values for the states are calculated from the end-effector to the base by backward recursion.

In this paper, the algorithm is implemented with reference to the work of Kim and Pollard \cite{kim2011fast} and modified with time scaling as {\bf Algorithm \ref{alg_recinvd}}. 

$ \bm{\xi}=(\bm{\omega},\bm{v}) \in {\mathbb{R}}^{6}, \bm{V}_i \in {\mathbb{R}}^{6} $ are generalized velocities (or twist) 
and $ \bm{\xi} $ is generalized velocity of joint frame at a reference frame,
$ \bm{F}_i \in {\mathbb{R}}^{6} $ are generalized forces,
$ J_i \in {\mathbb{R}}^{6 \times 6} $ is generalized inertia matrix,
$ \bm{g} \in {\mathbb{R}}^3 $ is a gravitational constant vector.
$ \hat{\cdot}:{\mathbb{R}}^{6} \rightarrow se(3)$ is defined as $ \hat{\bm{\xi}}= \begin{bmatrix} \hat{\bm{\omega}} & \bm{v} \\ \bm{0}^{\top} & 0 \\ \end{bmatrix} $
 where $ \hat{\bm{\omega}} = \begin{bmatrix} 0 & -{\omega}_3 & {\omega}_2 \\ {\omega}_3 & 0 & -{\omega}_1 \\ -{\omega}_2 & {\omega}_1 & 0 \end{bmatrix} $.
For $ se(3) $, an adjoint mapping $ Ad : SE(3) \times se(3) \rightarrow se(3) $ is defined as $ Ad_{G}(\hat{\bm{\xi}}) = G \hat{\bm{\xi}} G^{-1} $ ,
and a Lie bracket $ ad : se(3) \times se(3) \rightarrow se(3) $ is defined as $ ad_{\hat{\bm{\xi}}_1}(\hat{\bm{\xi}}_2) = \hat{\bm{\xi}}_1 \hat{\bm{\xi}}_2 - \hat{\bm{\xi}}_2 \hat{\bm{\xi}}_1 $ .
If $ \hat{\bm{\xi}} \in se(3) $ is represented as a vector $ \bm{\xi} \in {\mathbb{R}}^{6} $ , $ Ad $ and $ ad $  are represented in matrix form with using $ G = \begin{bmatrix} R & \bm{p} \\ \bm{0}^{\top} & 1 \\ \end{bmatrix} $ as
$ Ad_G(\bm{\xi}) = \begin{bmatrix} R & 0_{3 \times 3} \\ \hat{\bm{p}} R & R \\ \end{bmatrix} \bm{\xi} $,
$ ad_{\bm{\xi}_1}(\bm{\xi}_2) = \begin{bmatrix} \hat{\bm{\omega}}_1 & 0_{3 \times 3} \\ \hat{\bm{v}}_1 & \hat{\bm{\omega}}_1 \\ \end{bmatrix} \bm{\xi}_2 $ .
See details in the work of Lee et al. \cite{lee2005newton} and Park et al. \cite{park1995lie}.
\begin{algorithm}
\caption{Time Scaled Recursive Inverse Dynamics $\tilde{\bm{\tau}}(s)$ }
\label{alg_recinvd}
\begin{algorithmic}
\renewcommand{\algorithmicrequire}{\textbf{Input:}}
\renewcommand{\algorithmicensure}{\textbf{Output:}}
\Require  $ \tilde{\bm{q}}, \tilde{\bm{q}}^{\prime}, \tilde{\bm{q}}^{\prime\prime} \in {\mathbb{R}}^n , t_f \in {\mathbb{R}} $
\Ensure $ \tilde{\bm{\tau}} \in {\mathbb{R}}^n$
\State $ \bm{V}_0 \leftarrow \bm{0}, \;\; \bm{V}^{\prime}_0 \leftarrow \begin{pmatrix}\bm{0} \\ t_f \bm{g} \end{pmatrix}, \;\; \bm{F}_{n+1} \leftarrow \bm{0} $ \Comment{Initialization}
\For{$i \gets 1 \textrm{ to } n$} \Comment{Forward Recursion}
\State $ f_{i-1,i} = e^{\hat{\xi}_i \tilde{q}_i} $
\State $ \bm{V}_i = Ad_{f_{i-1,i}^{-1}} \bm{V}_{i-1} + \bm{\xi}_i \tilde{q}^{\prime}_i t_f^{-1} $
\State $ \bm{V}^{\prime}_i = Ad_{f_{i-1,i}^{-1}} \bm{V^{\prime}}_{i-1} + ad_{\bm{V}_i} \bm{\xi}_i \tilde{q}^{\prime}_i t_f^{-1} + \bm{\xi}_i \tilde{q}^{\prime\prime}_i t_f^{-2}$
\EndFor
\For{$i \gets n \textrm{ to } 1$} \Comment{Backward Recursion}
\State $ \bm{F}_i = J_i \bm{V}^{\prime}_i t_f^{-1} - ad_{\bm{V}_i}^{*}(J_i \bm{V}_i) + Ad_{f_{i,i+1}^{-1}}^{*}\bm{F}_{i+1} $
\State $ \tilde{\tau}_i = \bm{\xi}_i^{\top} \bm{F}_i $
\EndFor
\State \textbf{return} $ \tilde{\bm{\tau}} $
\end{algorithmic}
\end{algorithm}
%
\subsection{Efficient Inequality Constraint Checking Method for B-Spline Basis}\label{sec_bspline}
In our method, we can choose an arbitrary basis function such as a polynomial, orthogonal polynomial, wavelet, spline, etc.
If we adopt the B-Spline basis function, we can check semi-infinite linear inequality constraints without finite dense sampling.
We describe the efficient inequality checking method applicable for B-Spline.
In our proposed method, the B-spline order $ k \in \mathbb{N}_{+} $ and knot vector $ \bm{v} = (0=s_1, s_2, \cdots, s_{m+k}=1)^{\top} \in {\mathbb{R}}^{m+k} $ are given by users.
The B-spline basis of order $ k $ is as follows.
\begin{mydef}[B-Spline Basis of Order $ k $]
\begin{align}
& b_{i,1}(s) =
  \begin{cases}
    1 & \text{if $ s_i \leq s < s_{i+1} $} \\
    0 & \text{otherwise} 
  \end{cases} \\
& b_{i,k}(s) = \\
& \frac{s-s_i}{s_{i+k-1}-s_i}b_{i,k-1}(s) + \frac{s_{i+k}-s}{s_{i+k}-s_{i+1}}b_{i+1,k-1}(s)
\end{align}
\end{mydef}
The joint angular velocity is $ \tilde{\dot{\bm{q}}}(s) = A \bm{b}^{\prime}(s) t_f^{-1} $, where a derivative of B-spline basis $ b^{\prime}_i(s) $ is as follows \cite{de1978practical}.
\begin{multline}
{b}^{\prime}_{i,k}(s) = (k-1) \times \\
\left[ \frac{1}{s_{i+k-1}-s_i}b_{i,k-1}(s) - \frac{1}{s_{i+k}-s_{i+1}}b_{i+1,k-1}(s) \right]
\end{multline}
Since a B-spline has a convex hull property \cite{de1978practical}\cite{van2015optimal}, if the convex hull is in a feasible region, the feasible region always includes the B-Spline's trajectory.
Thus, the (finite number of) inequality constraints below is a sufficient condition for semi-infinite constraints (\ref{ineq_q})(\ref{ineq_qdot})\cite{van2015optimal}.
\begin{myprop}[Inequality of $ \tilde{\bm{q}}(s) = A \bm{b}(s) $ Linear Constraints for B-Spline Parameterization]
\begin{multline}
\underline{\bm{q}}\ \bm{1}_m^{\top} \ \underset{i,j}{\le} \ A \ \underset{i,j}{\le} \ \overline{\bm{q}}\ \bm{1}_m^{\top} \quad \Rightarrow \\
  \underline{\bm{q}} \le A \bm{b}(s) \le \overline{\bm{q}} \quad \forall s \in [0, 1] \label{eqn_q_ineq_ext}
\end{multline}
where $ \bm{1}_m = (\underbrace{1,\cdots,1}_{m})^{\top} \in {\mathbb{R}}^{m} $ .
\end{myprop}

\begin{myprop}[Inequality of $ \tilde{\bm{q}}^{\prime}(s) t_f = A \bm{b}^{\prime}(s) $ Linear Constraints for B-Spline Parameterization]
\begin{multline}
\frac{1}{k-1} t_f \underline{\dot{\bm{q}}} (\bm{v}_{k+1:k+m-1} - \bm{v}_{2:m})^{\top} \\
 \ \underset{i,j}{\le} \  A \left( \begin{pmatrix} I_{m-1 \times m-1} \\ \bm{0}_{m-1}^{\top} \end{pmatrix} - \begin{pmatrix} \bm{0}_{m-1}^{\top} \\ I_{m-1 \times m-1} \end{pmatrix} \right) \\
 \ \underset{i,j}{\le} \  \frac{1}{k-1} t_f \overline{\dot{\bm{q}}} (\bm{v}_{k+1:k+m-1} - \bm{v}_{2:m})^{\top} \\ 
  \Rightarrow \quad
    \underline{\dot{\bm{q}}} \le A \dot{\bm{b}}(s) \le \overline{\dot{\bm{q}}} \quad \forall s \in [0,1] \label{eqn_qdot_ineq_ext}
\end{multline}
\end{myprop}
where
  $ \underset{i,j}{\le} $ is elementwise inequality of a matrix,
  $ \bm{v}_{i:j} $ means a vector part of indices $ \{i, \cdots , j\} $ \footnote{Matlab like notation}, 
  $ I_{m-1 \times m-1} \in {\mathbb{R}}^{m-1 \times m-1} $ is an identity matrix,
  and $ \bm{0}_{m-1} = (\underbrace{0,\cdots,0}_{m-1})^{\top} \in {\mathbb{R}}^{m-1} $ .
Semi-infinite inequality for the whole continuous range $ [0,1] $ can be checked by using the above linear inequality of $ A $ .
Note that the above property is a sufficient condition for checking (\ref{ineq_q})(\ref{ineq_qdot}). Thus, this method could produce conservative constraints \cite{van2015optimal}.
%
\section{Numerical Experiments}
We conducted numerical experiment to assess whether our algorithm can be calculated correctly with practical computational time and generate an appropriate solution.
First, we evaluated basic behavior by applying the algorithm to a 2-link manipulator.
We next applied the algorithm to a practical 6-DoF manipulator that has many links to assess the scalability of our algorithm.
The algorithm is implemented by Matlab and optimization problems are solved by fmincon with SQP.
The Derivative of torque constraints, which is a very complicated expression, is calculated by using an automatic differentiation \cite{griewank2008evaluating} module we made.
We checked correctness of inverse dynamics implementation and visualized result motions by using simscape multibody.
%
\subsection{In the case of a 2-Link Manipulator}
\begin{figure*}
\centering
\vspace*{-2cm}
\includegraphics[width=1.0\textwidth, bb=9 79 780 501]{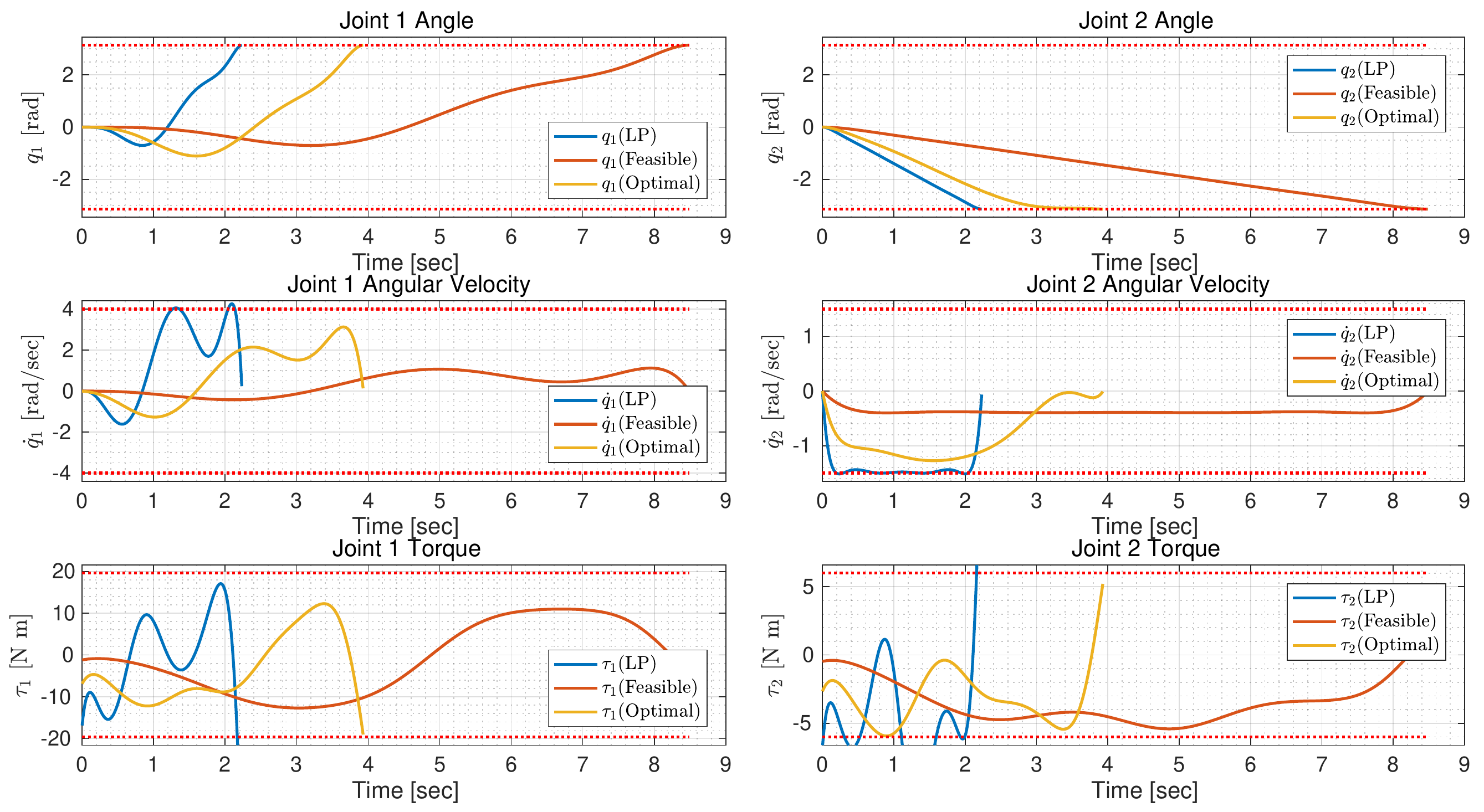}
\vspace*{+1.2cm}
\caption{Trajectory of 2-Link Manipulator Planning. Dotted lines are upper and lower bounds of the state or control. Three trajectories in each graph correspond to each step in {\bf Algorithm \ref{alg_feas_init}}. First, {\bf solution LP} satisfies state (angle and angular velocity) constrains. Next, {\bf solution feas} satisfies torque constraints by extending $ t_f $ . Lastly, {\bf solution opt} lowers cost $ t_f $ without violating any constraints.}
\label{fig_2link_init}
\end{figure*}

\begin{figure*}
\begin{minipage}{0.33\hsize}
\centering
\includegraphics[scale=0.125, bb=0 0 1254 710]{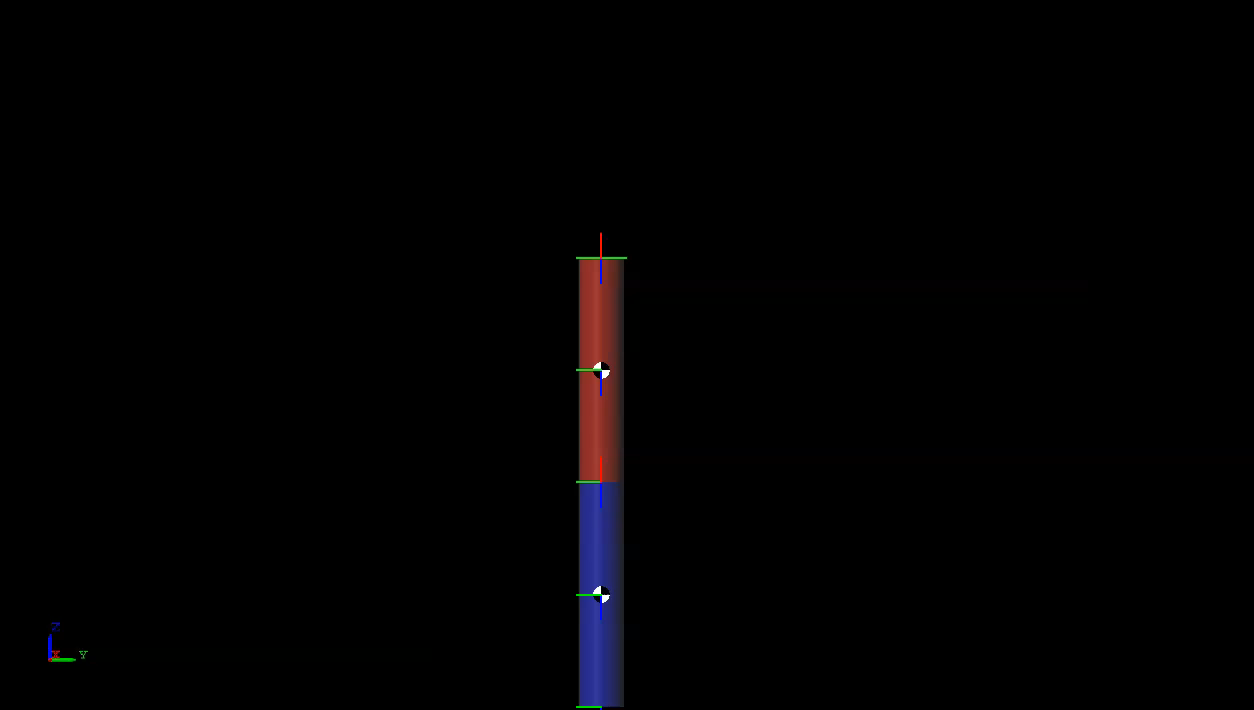}
(a)Initial state
\end{minipage}
\begin{minipage}{0.33\hsize}
\centering
\includegraphics[scale=0.125, bb=0 0 1254 710]{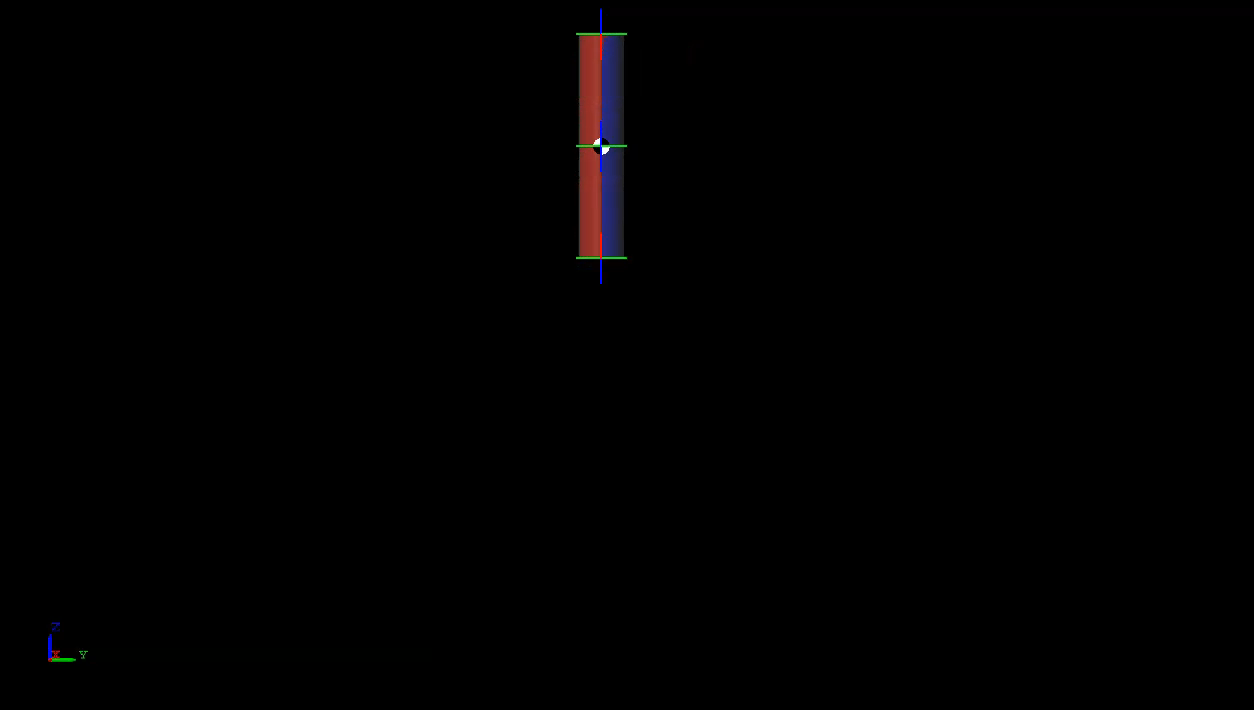}
(b)Final state
\end{minipage}
\begin{minipage}{0.33\hsize}
\centering
\includegraphics[scale=0.125, bb=0 0 1254 710]{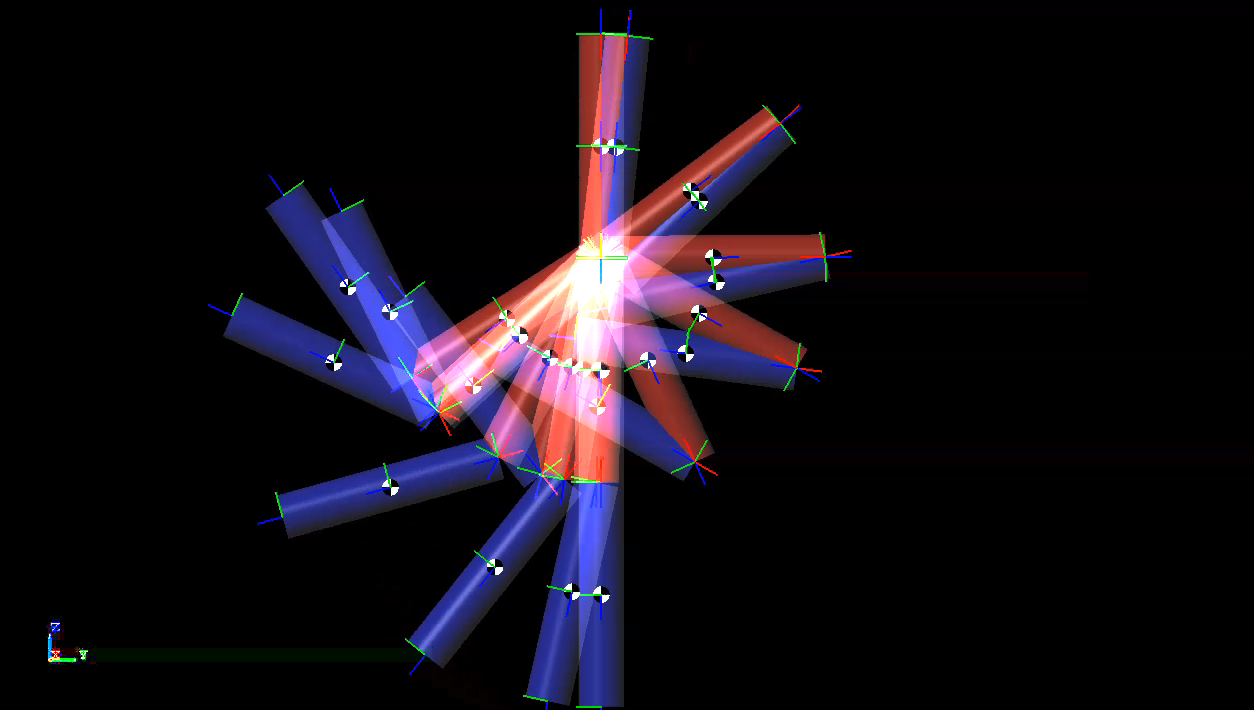}
(c)Planned motion
\end{minipage}
\caption{Planned motion of 2-link manipulator}
\label{fig_2link_animation}
\end{figure*}

Initial states, final states, and the planned motion result are shown in Fig. \ref{fig_2link_animation}.
Problem settings are as follows.
\begin{itemize}
\item Cost function $ J(A,t_f) = t_f $, thus the problem is a time-optimal set-point problem.
\item State and control constraints are
  \begin{align*}
     \bm{q}_0                 &= (0,0),        & \dot{\bm{q}}_0          &= (0,0), \\
     \bm{q}_f                 &= (\pi,-\pi),   & \dot{\bm{q}}_f          &= (0,0), \\
     \underline{\bm{q}}       &= (-\pi,-\pi),  & \overline{\bm{q}}       &= (\pi,\pi), \\
     \underline{\dot{\bm{q}}} &= (-4.0,-1.5),  & \overline{\dot{\bm{q}}} &= (4.0,1.5), \\
     \underline{\bm{\tau}}    &= (-19.6,-6.0), & \overline{\bm{\tau}}    &= (19.6,6.0)
  \end{align*}
  (torque limit satisfies the assumption of enough force for the gravity term)
\item Basis function is a polynomial basis, and order is 10.
\item The number of sample points $ N $ for checking inequality is 24.
\end{itemize}
The planned trajectory details are shown in Fig.\ref{fig_2link_init}.
The implemented algorithm worked correctly, and observed behaviors were as expected.
\begin{itemize}
\item {\bf Solution LP} satisfies all state constraints but violates torque constraints.
\item {\bf Solution feas} satisfies all state and torque constraints and is a feasible but not optimal solution.
\item {\bf Solution opt} satisfies all state and torque constraints and lowers cost $ t_f $.
\end{itemize}

We actually made torque constraints slightly stricter in step 2 of the algorithm because we observed some cases in which if {\bf solution feas} is very close to the torque constraints, the NLP algorithm can barely improve the solution.

%
%
\subsection{In the case of a 6-DoF Manipulator}
\begin{figure}
\centering
\vspace*{-0.5cm}
\includegraphics[height=50mm, bb=85 26 561 506]{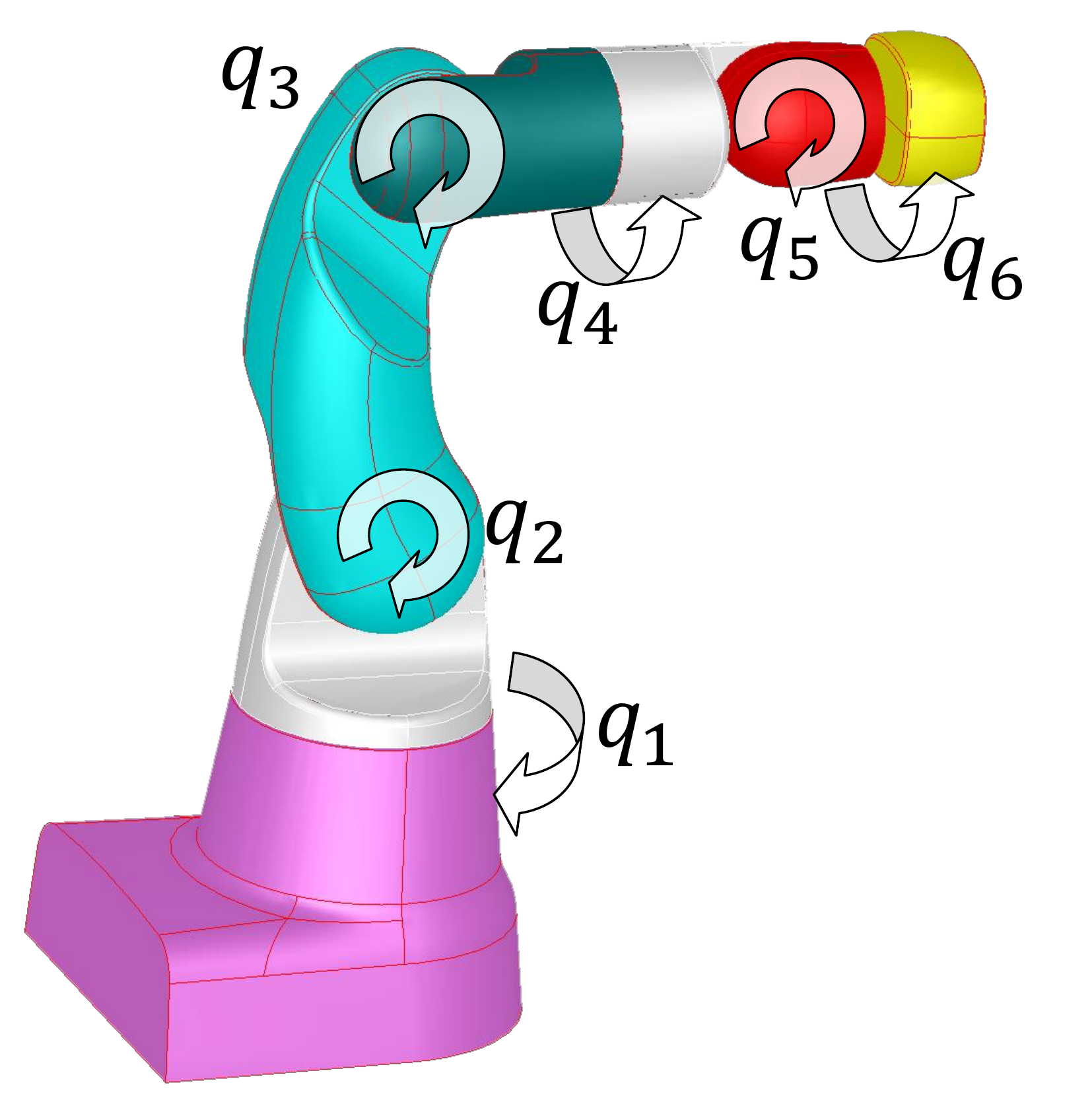} 
\vspace*{+0.5cm}
\caption{CAD image of real 6-DoF robot manipulator for our numerical experiments}
\label{fig_robot}
\vspace{-1cm} 
\end{figure}

\begin{figure*}
\begin{minipage}{0.33\hsize}
\centering
\includegraphics[scale=0.13, bb=0 0 1230 854]{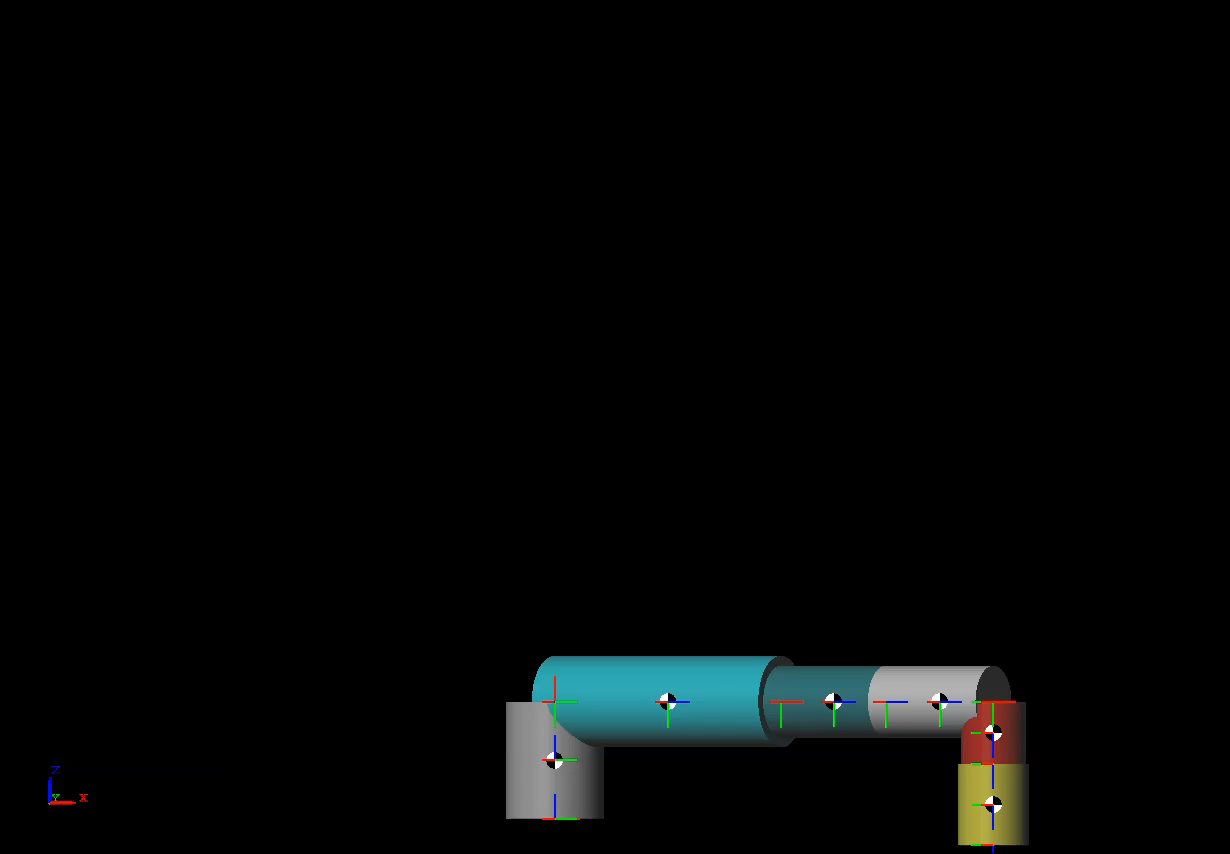} 
(a) Initial state
\end{minipage}
\begin{minipage}{0.33\hsize}
\centering
\includegraphics[scale=0.13, bb=0 0 1230 854]{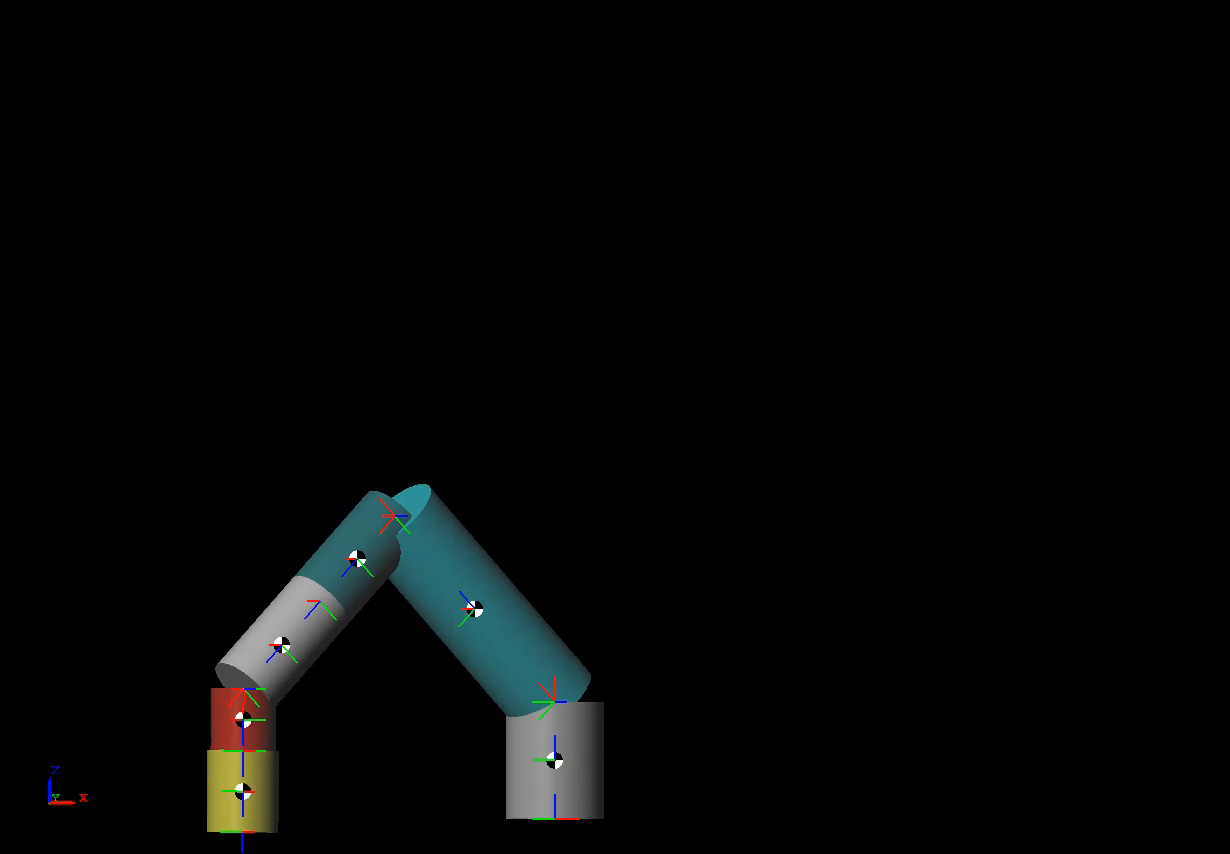} 
(b) Final state
\end{minipage}
\begin{minipage}{0.33\hsize}
\centering
\includegraphics[scale=0.13, bb=0 0 1230 854]{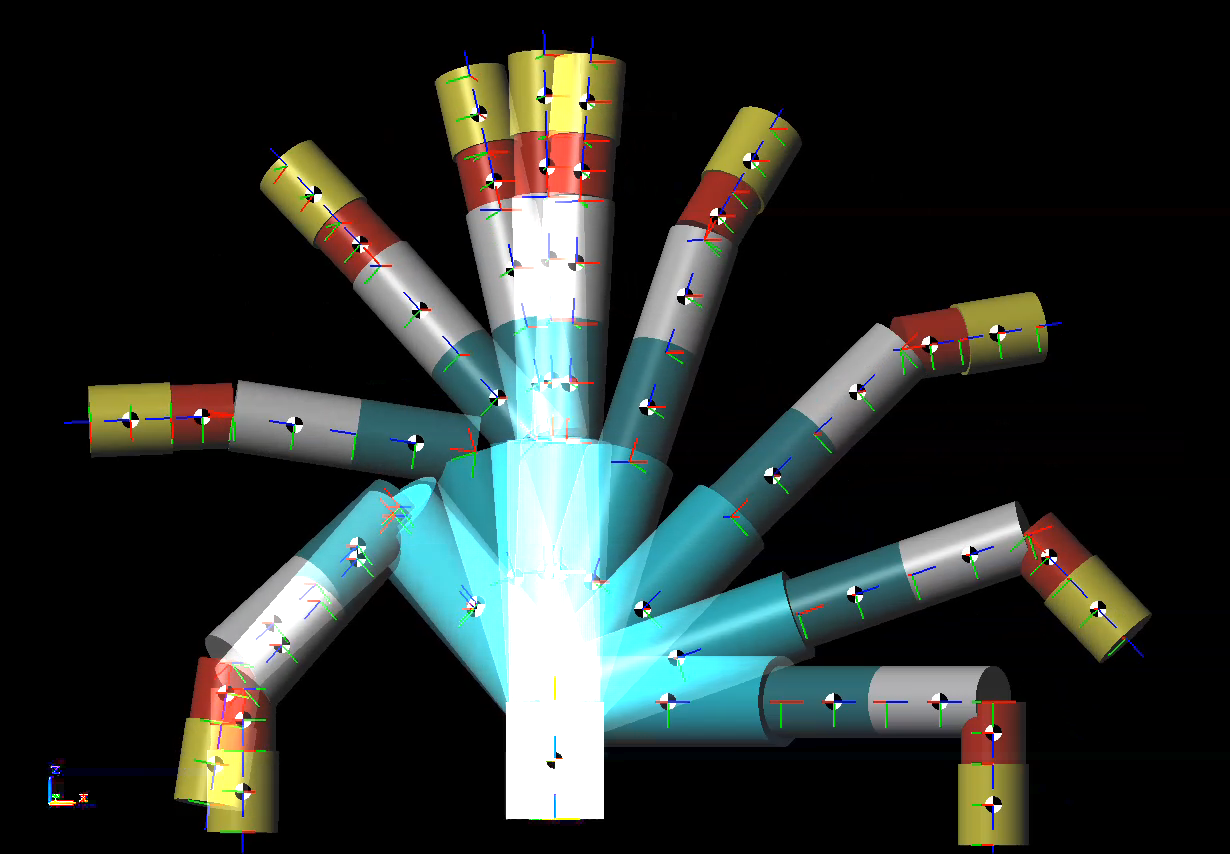} 
(c) Planned motion
\end{minipage}
\caption{Planned motion of 6-DoF manipulator}
\label{fig_6dof_animation}
\end{figure*}

\begin{figure}
\centering
\vspace*{-2cm}
\includegraphics[scale=0.45, bb=39 170 556 672]{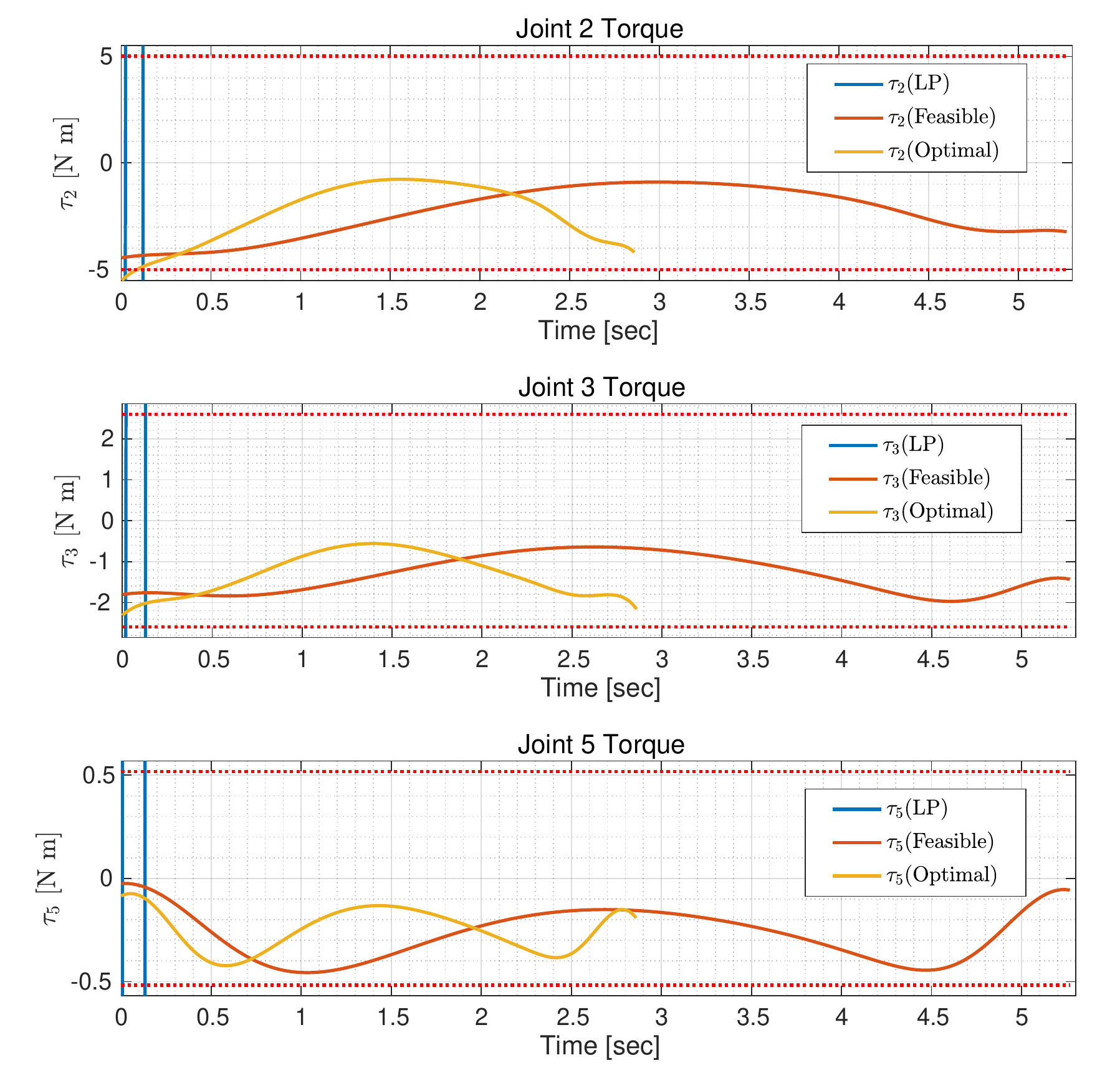}
\vspace*{+2cm}
\caption{Planned torque trajectories of joints 2,3 and 5}
\label{fig_6dof_trajectory}
\end{figure}

Initial states, final states, and the planned motion result are shown in Fig. \ref{fig_6dof_animation}.
The problem settings are as follows.
\begin{itemize}
\item Cost function $ J(A,t_f) = t_f $
\item The manipulator has 6-links and 6-DoFs as shown in Fig. \ref{fig_robot}
\item Basis function is B-spline, order is $ k=9 $, and the number of knots is $ 24 $
\end{itemize}
\vspace*{+0.5cm}
%
Both trajectories of the states and controls of the result successfully satisfied the inequality constraints for them.
We show some torque trajectories that move largely in Fig.\ref{fig_6dof_trajectory}.
%
%
%
\section{Discussion}
We found that our method can correctly construct feasible initial points and prevent failure to find a feasible solution with NLP.
The difference between our method and using multiple heuristic initial points is its reliability: our method can find a feasible solution for our non-convex problem without fail.

However, there are several problems that our method does not solve.

The first problem is that improved {\bf solution opt} is local minima, and how much global optimum and {\bf solution opt} differ is not clear.
An important research direction is to find solutions nearer to the global optimum.
For example, one idea is apply a tight convex relaxation to torque expression.

The second unsolved problem is which basis function is best for our problem.
We tried polynomial basis, B-spline basis, and orthogonal polynomial basis (e.g. Chebyshev polynomial, Legendre polynomial). In our experiments, polynomial and B-Spline bases generate better results, but their mechanisms are not clear.
An interesting direction is to use another type of basis such as wavelet spline that has both short and wide region base.

The third problem is that our algorithm correctly works under the assumption in {\bf Property \ref{prop_torque}}, but we do not know whether it works correctly under harder constraints.
Thus, the assumption should be removed and the algorithm should be applied under harder constraints.
In our preliminary experiment under harder torque constraints, we observed an energy storing motion that first uses back swing to store motion and next achieves a final state that cannot be reached directly. Researching smart motions under harder constraints is an interesting direction.
\section{Conclusion and Future Work}
We presented a method for unfailingly finding a feasible solution for a flatness-based robot manipulator motion planning problem under state and control constraints when it is a non-convex optimization problem.
Our method can reliably find a feasible solution under practical assumptions even though common methods that initialize multiple infeasible initial points and use general nonlinear programming (NLP) often fail and cannot reliably find a feasible solution.

Our method first finds a solution that satisfies state constraints, transforms it to satisfy control constraints, and lowers its cost.
We found that our method correctly works for a multi-link robot manipulator that has two links or six-degrees of freedom (DoFs).
Future directions are to find the best basis function, find a solution nearer to the global optimum, and construct a reliable method under constraints harder than those assumed in this paper.
%
\section{Acknowledgments}
The authors gratefully acknowledge the contributions of the reviewers.
\bibliographystyle{plain}
\bibliography{SICE_ISCS2017}

\end{document}